\documentclass[twocolumn,english,superscriptaddress,citeautoscript,showpacs,preprintnumbers,amsmath,amssymb,prl,floatfix,footinbib]{revtex4}
\usepackage[utf8x]{inputenc}
\usepackage[scaled=2]{helvet}
\usepackage[T1]{fontenc}
\usepackage[utf8]{luainputenc}
\usepackage{multirow}
\usepackage{textcomp}
\usepackage{amsmath}
\usepackage{amssymb}
\usepackage{graphicx}
\usepackage{subscript}
\usepackage{epstopdf}
\makeatletter

\newcommand{\lyxmathsym}[1]{\ifmmode\begingroup\def\b@ld{bold}
  \text{\ifx\math@version\b@ld\bfseries\fi#1}\endgroup\else#1\fi}

\@ifundefined{textcolor}{}
{%
 \definecolor{BLACK}{gray}{0}
 \definecolor{WHITE}{gray}{1}
 \definecolor{RED}{rgb}{1,0,0}
 \definecolor{GREEN}{rgb}{0,1,0}
 \definecolor{BLUE}{rgb}{0,0,1}
 \definecolor{CYAN}{cmyk}{1,0,0,0}
 \definecolor{MAGENTA}{cmyk}{0,1,0,0}
 \definecolor{YELLOW}{cmyk}{0,0,1,0}
}

\usepackage[scaled=2]{helvet}\usepackage{amstext}

\makeatletter
\@ifundefined{textcolor}{}{\definecolor{BLACK}{gray}{0}\definecolor{WHITE}{gray}{1}\definecolor{RED}{rgb}{1,0,0}\definecolor{GREEN}{rgb}{0,1,0}\definecolor{BLUE}{rgb}{0,0,1}\definecolor{CYAN}{cmyk}{1,0,0,0}\definecolor{MAGENTA}{cmyk}{0,1,0,0}\definecolor{YELLOW}{cmyk}{0,0,1,0}}

\usepackage{dcolumn}
\usepackage{bm}
\usepackage{times}
\usepackage{hyperref}
\hypersetup{colorlinks=true,linkcolor=red,citecolor=blue}
\makeatother

\usepackage{babel}

\makeatother
\usepackage{babel}

\begin{document}


\title{Non-collinear magnetic structure and anisotropic magnetoelastic coupling
\\ in cobalt pyrovanadate Co$_2$V$_2$O$_7$ $^\dag$}

\author{W. H. Ji}
\affiliation{J\"{u}lich Centre for Neutron Science JCNS and Peter Gr\"{u}nberg Institut PGI, JARA-FIT, Forschungszentrum J\"{u}lich GmbH, 52425 J\"{u}lich, Germany}

\author{Y. C. Sun}
\affiliation{Wuhan National High Magnetic Field Center, Huazhong University of Science and Technology, Wuhan 430074, P. R. China}

\author{C. M. N. Kumar}
\affiliation{Institut f\"{u}r Festk\"{o}rperphysik, TU Wien, Wiedner Hauptstr. 8-10/138, 1040 Wien, Austria}

\author{C. Li}
\affiliation{J\"{u}lich Centre for Neutron Science JCNS, Forschungszentrum J\"{u}lich GmbH, Outstation at SNS, POB 2008, 1 Bethel Valley Rd. Oak Ridge, TN 37831-6473, USA}

\author{S. Nandi}
\affiliation{J\"{u}lich Centre for Neutron Science JCNS and Peter Gr\"{u}nberg Institut PGI, JARA-FIT, Forschungszentrum J\"{u}lich GmbH, 52425 J\"{u}lich, Germany}

\author{W. T. Jin}
\affiliation{J\"{u}lich Centre for Neutron Science JCNS at Heinz Maier-Leibnitz Zentrum, Forschungszentrum J\"{u}lich GmbH, Lichtenbergstra{\ss}e 1, 85747 Garching, Germany}

\author{Y. Su}
\affiliation{J\"{u}lich Centre for Neutron Science JCNS at Heinz Maier-Leibnitz Zentrum, Forschungszentrum J\"{u}lich GmbH, Lichtenbergstra{\ss}e 1, 85747 Garching, Germany}

\author{X. Sun}
\affiliation{J\"{u}lich Centre for Neutron Science JCNS and Peter Gr\"{u}nberg Institut PGI, JARA-FIT, Forschungszentrum J\"{u}lich GmbH, 52425 J\"{u}lich, Germany}

\author{Y. Lee}
\affiliation{Ames Laboratory, U.S. Department of Energy, Ames, Iowa 50011}

\author{B. Harmon}
\affiliation{Ames Laboratory, U.S. Department of Energy, Ames, Iowa 50011}

\author{L. Ke}
\email{liqinke@ameslab.gov}
\affiliation{Ames Laboratory, U.S. Department of Energy, Ames, Iowa 50011}

\author{Z. W. Ouyang}
\email{zwouyang@mail.hust.edu.cn}
\affiliation{Wuhan National High Magnetic Field Center, Huazhong University of Science and Technology, Wuhan 430074, P. R. China}

\author{Y. Xiao}
\email{xiaoyg@pkusz.edu.cn}
\affiliation{School of Advanced Materials, Peking University Shenzhen Graduate School, Shenzhen 518055, China}

\author{Th. Br\"{u}ckel}
\affiliation{J\"{u}lich Centre for Neutron Science JCNS and Peter Gr\"{u}nberg Institut PGI, JARA-FIT, Forschungszentrum J\"{u}lich GmbH, 52425 J\"{u}lich, Germany}

\date{\today}

\begin{abstract}

The Co$_2$V$_2$O$_7$ is recently reported to exhibit amazing magnetic field-induced magnetization plateaus and ferroelectricity, but its magnetic ground state remains ambiguous due to its structural complexity. Magnetometry measurements, and time-of-flight neutron powder diffraction (NPD) have been employed to study the structural and magnetic properties of Co$_2$V$_2$O$_7$, which consists of two non-equivalent Co sites. Upon cooling below the Ne\'{e}l temperature T$_N$ = ~6.3 K, we observe magnetic Bragg peaks at 2K in NPD which indicated the formation of long range magnetic order of Co$^{2+}$ moments. After symmetry analysis and magnetic structure refinement, we demonstrate that Co$_2$V$_2$O$_7$ possesses a complicated non-collinear magnetic ground state with Co moments mainly located in \emph{b-c} plane and forming a non-collinear spin-chain-like structure along the \emph{c}-axis. The ab initio calculations demonstrate that the non-collinear magnetic structure is more stable than various ferromagnetic states at low temperature. The non-collinear magnetic structure with canted $\uparrow$$\uparrow$$\downarrow$$\downarrow$ spin configuration is considered as the origin of magnetoelectric coupling in Co$_2$V$_2$O$_7$ because the inequivalent exchange striction induced by the spin-exchange interaction between the neighboring spins is the driving force of ferroelectricity. Besides, it is found that the deviation of lattice parameters \emph{a} and \emph{b} is opposite below T$_N$, while the lattice parameter \emph{c} and $\beta$ stay almost constant below T$_N$, evidencing the anisotropic magnetoelastic coupling in Co$_2$V$_2$O$_7$.

\end{abstract}

\pacs{74.70.Xa, 74.62.-c, 75.47.-m, 75.25.Dk}
\maketitle

\section{Introduction}

Low-dimensional spin systems have attracted extensive interest as their low dimensionality and complex multiple-spin exchange interactions often lead to novel ground states as well as intriguing magnetic behavior, such as spinon excitations in one-dimensional (1D) spin chain system KCuF$_3$ \cite{Tennant1,Lake1}, and a spin liquid state in two-dimensional triangular lattice antiferromagnet NiGa$_2$S$_4$ \cite{Nakatsuji1697}. The complex magnetic and structural characteristics of low-dimensional spin systems will also make its magnetic state highly sensitive to external perturbations including temperature and magnetic field. For instance, the field-induced quantum phase transition from the N\'{e}el ordered phase to the spin liquid one is observed in quasi-1D antiferromagnet BaCo$_2$V$_2$O$_8$ \cite{Kimura1}, and a cascade of magntization plateaus in spin chain system Ca$_3$Co$_2$O$_6$ \cite{Lampen,Kageyama,Kudasov}.

The family of cobalt vanadium oxides Co$_x$V$_2$O$_{5+x}$ with \emph{x} = (1, 2 or 3) has attracted special attention due to their complex crystal structure and magnetic behavior, depending on the value of the subscript \emph{x} \cite{HeJACS,Drees,HeJSSC2009,Balakrishnan}. The \emph{x} = 1 compound CoV$_2$O$_6$ is a low-dimensional spin system which crystalizes in two structural polymorphs with one having a monoclinic unit cell and the other having a triclinic unit cell \cite{MOCALA1987299}, i.e. $\alpha$-CoV$_2$O$_6$ and $\gamma$-CoV$_2$O$_6$. Although two polymorphs possess different structural symmetries, both $\alpha$-CoV$_2$O$_6$ and $\gamma$-CoV$_2$O$_6$ are quasi-1D Ising spin-chain compounds. Application of external magnetic field leads to the observation of unusual 1/3 magnetization plateaus in both polymorphs \cite{HeJACS,LenertzJPCC,NandiJAP,Shu2016129}, which is attributed to the strong spin exchange and the strong coupling between orbital, magnetic and structural orders \cite{WallingtonPRB,HollmannPRB,KimberPRB}. For \emph{x} = 3, however, the Co$_3$V$_2$O$_8$ consists of buckled Kagome layers of edge-sharing CoO$_6$ octahedra separated by non-magnetic V$_5$O$_4$ tetrahedra, showing a geometrically frustrated lattice and a strong two-dimensional magnetic character \cite{ChenPRB,RamazanogluPRB}. Interestingly, a dielectric anomaly is observed around the ferromagnetic (FM) transition temperature T$_C$ indicating the existence of strong coupling between spin and charges in Co$_3$V$_2$O$_8$ \cite{BellidoJPCM}.

When \emph{x} = 2, Co$_2$V$_2$O$_7$ adopts a dichromate K$_2$Cr$_2$O$_7$-type structure, which crystallizes in a monoclinic system with space group $P2_1/c$ \cite{Sauerbrei}. As shown in Fig. 1, Co$_2$V$_2$O$_7$ possesses two different Co$^{2+}$ sites with two differing local bonding geometries, which are linked in a geometry that can be referred to as a zigzag chain consisting of arrays of edge-shared CoO$_6$ octahedra. The zigzag chains are interspersed with nonmagnetic V$^{5+}$ in VO$_4$ pyramid blocks, resulting in a quasi-1D structural arrangement \cite{HeJSSC2009}. Touaiher \emph{et al.} \cite{touaiher2004crystal} reported that Co$_2$V$_2$O$_7$ shows a ferromagnetic ordering while Ni$_2$V$_2$O$_7$ shows an antiferromagnetic (AFM) ordering. However, He \emph{et al.} \cite{he2009magnetic} argued that Co$_2$V$_2$O$_7$ and Ni$_2$V$_2$O$_7$ are likely possessing a similar magnetic ordering since they have the same crystal structure. Recently, two fascinating magnetization plateaus–a 1/2-like plateau and a 3/4 plateau, are observed in Co$_2$V$_2$O$_7$, depending on crystallographic directions \cite{Ouyang1,Ouyang2}, which is quite different from the isostructural Ni$_2$V$_2$O$_7$ with a 1/2 plateau \cite{Ouyang3}. The first-principles calculations show that the AFM ground state results in a $\uparrow$$\downarrow$$\uparrow$$\downarrow$-type of spin structure under the assumption that the spin configurations are collinear in the calculations \cite{Ouyang1}. In addition, magnetic field induced ferroelectricity was reported and found to be correlated with the magnetization plateaus in both Co$_2$V$_2$O$_7$ and Ni$_2$V$_2$O$_7$ \cite{Ouyang2}. Moreover, the deviations of the dielectric constant with respect to lattice contributions is evidently observed in Co$_2$V$_2$O$_7$ under zero magnetic field, suggesting the appearance of magnetostrictive effects induced by magnetic interactions \cite{SanchezJAP}. Thus, by investigating the magnetic ground state and the response of the lattice to the magnetic phase transitions, it is possible to develop a deeper understanding of the spin-lattice couplings in Co$_2$V$_2$O$_7$ \cite{chen2018magnetic}.

Although there are a few reports on Co$_2$V$_2$O$_7$, its magnetic structure remains ambiguous because of the complexity of the structure, which is crucial to understand the field-induced magnetization plateaus, ferroelectricity as well as the magnetostrictive effect. In this paper, we report the structural and magnetic properties of Co$_2$V$_2$O$_7$ by magnetization and neutron powder diffraction methods. Both nuclear and magnetic structures are determined through a refinement of high resolution neutron powder diffraction data at low temperature of 2 K. It is found that Co$_2$V$_2$O$_7$ exhibits a non-collinear magnetic structure with Co spins forming a canted $\uparrow$$\uparrow$$\downarrow$$\downarrow$ spin chain. The non-collinear magnetic ground state and electronic structure of Co$_2$V$_2$O$_7$ can be well understood based on ab initio calculations. Moreover, according to the exchange striction model, the appearance of electric polarization can be predicted in Co$_2$V$_2$O$_7$ along the chain direction.

\section{Experimental Methods}

\subsection{Sample preparation and characterisation}
A Polycrystalline sample of Co$_2$V$_2$O$_7$ was prepared using the sol-gel method by dissolving stoichiometric amounts of high purity Co(NO$_3$)$_2$ $\cdot$ 6H$_2$O and NH$_4$VO$_3$ in deionized water. The X-ray powder diffraction measurement was carried out on a Huber diffractometer with Cu K$\alpha$ radiation $\lambda$ = 1.5406 {\AA} and a graphite monochromator. A Quantum Design Physical Property Measurement System (PPMS) was used to characterize the magnetic properties of the Co$_2$V$_2$O$_7$ powder sample in the temperature range from 2 to 350 K and the magnetic field range from 0 to 9 T. The temperature dependence of magnetization was measured during warming from 2 to 350 K under 1000 Oe after cooling down from 350 K without the applied field (zero field cooled-ZFC) or under the applied field (field cooled-FC), while the field dependence of magnetization was measured in magnetic field up to 9 T at selective temperatures.
\begin{figure}[h]
	\centering
	\includegraphics[height=8.5cm]{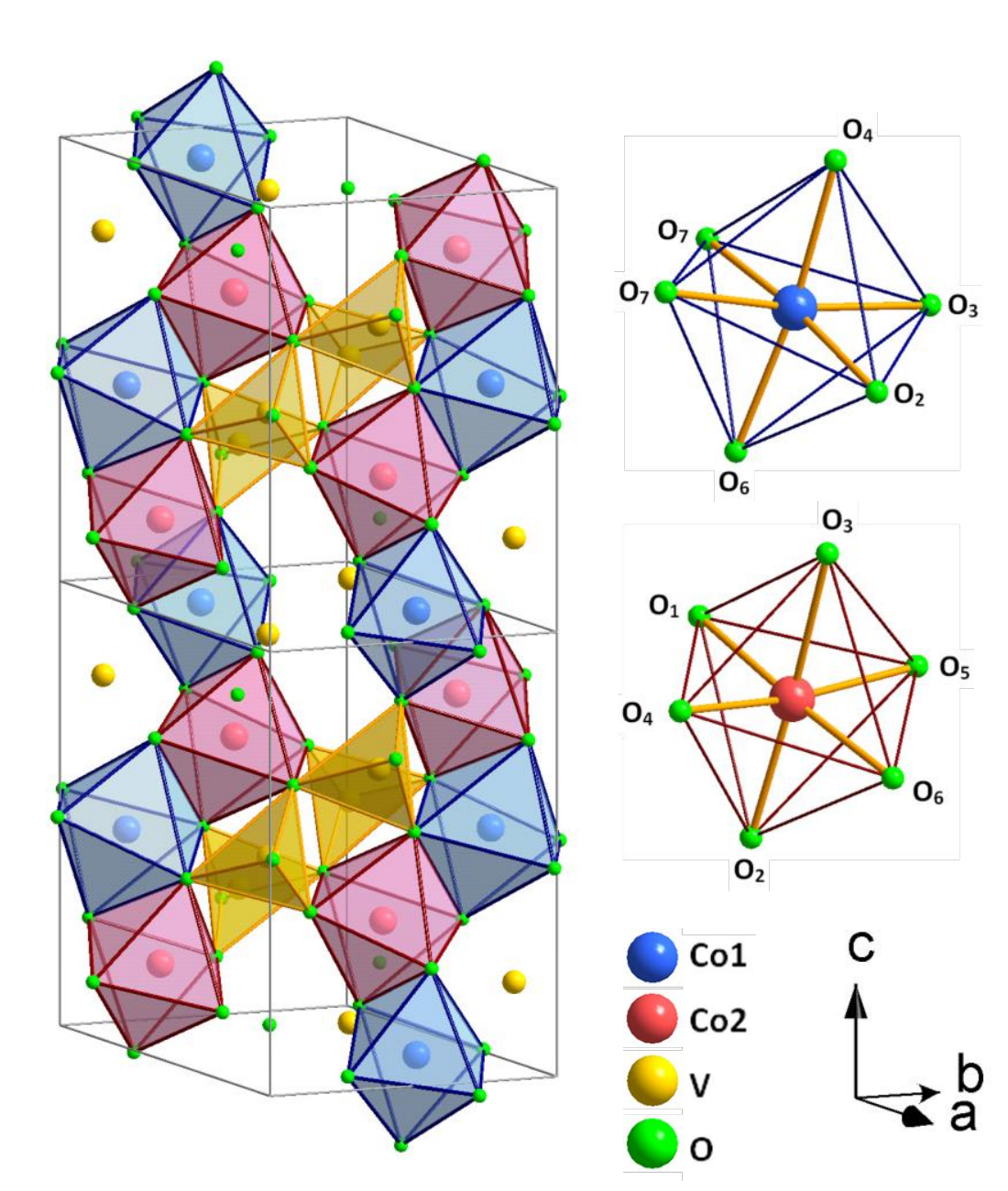}
	\caption{The crystal structure of Co$_2$V$_2$O$_7$ at 300 K, which has the prototypical K$_2$Cr$_2$O$_7$-type structure. The crystal structure can be described as the chains of edge-sharing CoO$_6$ octahedra with VO$_4$ tetrahedra in between. Two different types of CoO$_6$ octahedra, i.e. Co$_1$O$_6$ and Co$_2$O$_6$, are differentiated by color. The Co-O bond distances of Co$_1$O$_6$ and Co$_2$O$_6$ are illustrated for comparison.}
	\label{fgr:example}
\end{figure}

To study the crystal and magnetic structure of Co$_2$V$_2$O$_7$ systematically, time-of-flight (TOF) neutron powder diffraction (NPD) experiments were performed on the high resolution neutron powder diffractometer POWGEN \cite{huq2011powgen} at the Spallation Neutron Source (SNS) in Oak Ridge National Laboratory. About 5 gram of powder sample was loaded in an 8 mm diameter vanadium sample holder and then installed in the helium cryostat that can reach low temperatures down to 2 K. Neutron diffraction data were collected for the Co$_2$V$_2$O$_7$ powder sample in the temperature range from 2 to 300 K. For each temperature the data were collected using two different center wavelengths, viz., 1.066 and 3.731 {\AA} to cover a large d-space range of 0.5-12 angstrom. The representation analysis were carried out using the program SARAh \cite{wills2000new}, whereas the crystal and magnetic structure refinements were carried out from the NPD data using the Rietveld refinement program FullProf \cite{rodriguez1993recent}.

\subsection{Calculation details}
We have calculated electronic structure, magnetic properties and total energies difference between FM, AFM, and the experimental non-collinear (NC) magnetic states using the Vienna ab initio simulation package (VASP) \cite{kresse1993ab,kresse1996g}. The nuclei and core electrons were described by the projector augmented wave (PAW) potential \cite{kresse1999ultrasoft}. The wave functions of valence electrons were expanded in a plane-wave basis set with cutoff energy of up to 520 eV. The VASP calculations were performed with 240 \emph{k} points in IBZ. The correlation effects are also considered by using the generalized gradient approximation (GGA)+\emph{U} method \cite{dudarev1998electron}. For transition metals, typical  \emph{U} values are between 1.5 to 6 eV \cite{csacsiouglu2011effective}. Here, various \emph{U} parameters between 0 to 4.8 eV are used to investigate the dependence of spin configuration on correlation effects. The experimental lattice parameters and atomic coordinates measured through neutron diffraction at 2K, as listed in Table I, were used in all calculations.

\section{Experimental results and discussion}

\subsection{Magnetization and magnetic phase transition}
\begin{figure}[h]
	\centering
	\includegraphics[height=5.6cm]{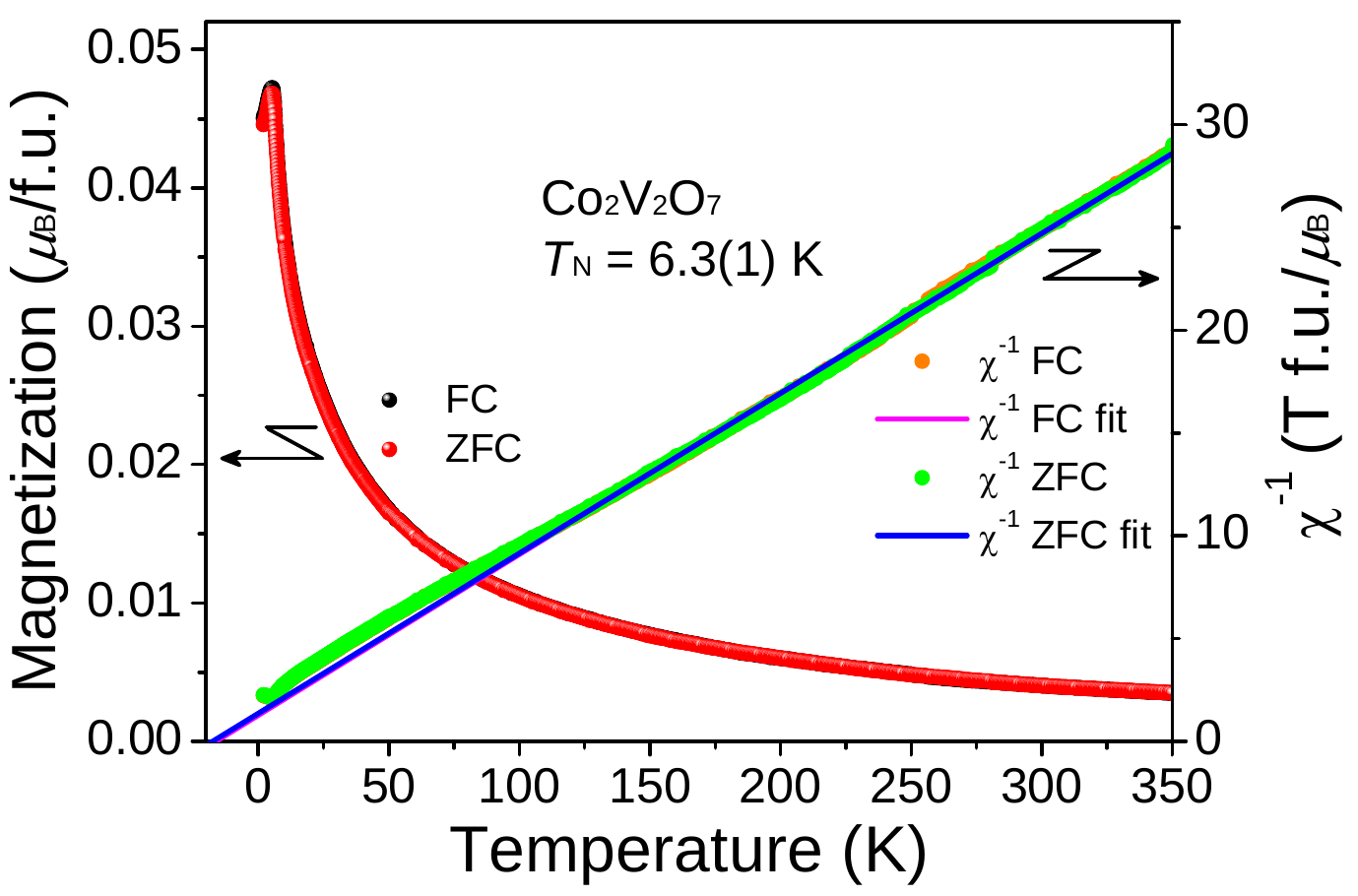}
	\caption{ Temperature dependence of the magnetization and inverse magnetic susceptibility $\chi$$^{-1}$ of Co$_2$V$_2$O$_7$ measured in zero-field-cooling and field-cooling mode under magnetic field of 1000 Oe. Solid lines indicate Curie-Weiss fits to inverse susceptibility in high temperature region from 200 to 350 K as described in the text.}
	\label{fgr:example}
\end{figure}

Figure 2 shows the temperature dependence of the magnetization of the Co$_2$V$_2$O$_7$ polycrystalline sample under magnetic fields of 1000 Oe. With decreasing temperature, the magnetization increases smoothly in a temperature range from 350 K to 6.3 K, while it starts decreasing when the sample is cooled below 6.3 K. The phase transition at 6.3(1) K can be clearly observed as indicated by the maximum of the magnetization data. The temperature of 6.3(1) K can be denoted as the N\'{e}el temperature \emph{T}$_N$ where the paramagnetic state develops from the antiferromagnetic phase with increasing temperature, as will be demonstrated by neutron diffraction results in the following text.

Temperature dependence of the inverse susceptibility 1/$\chi$ of Co$_2$V$_2$O$_7$ compound is also exhibited in Fig. 2. It is known that the magnetic susceptibility of localized noninteracting magnetic ions in high temperature region can be written as: $\chi(T) = \frac{C}{T - \Theta_{CW}}$, where \emph{C} and $\theta_{CW}$ are the Curie constant and Curie-Weiss temperature, respectively. The Curie constant can be expressed as $C = N M^2_{eff} \mu^2_B/3k_B$, where N is the density of magnetic ions, $M_{eff}$ is the effective paramagnetic moment, $k_B$ is the Boltzmann constant, and $\mu_B$ is the Bohr magneton. As indicated in Fig. 2, the susceptibility strictly follows the Curie-Weiss behavior in high temperature range, while the inverse susceptibility 1/$\chi$ deviated from the Curie-Weiss estimation below 120 K, indicating the onset of considerable magnetic correlations. The effective paramagnetic moment is deduced to be 3.77(1) $\mu_B \,$ and it matches very well with the effective magnetic moment of Co$^{2+}$, which can be obtained using $p_{eff} = g \sqrt{S(S+1)}$, where $g$ = 2 is the Land\'{e} \emph{g} factor and \emph{S} = 3/2. The fit of high temperature susceptibility by means of Curie-Weiss law gives a negative Curie-Weiss temperature of -16.1(5) K, indicating that the dominant magnetic interactions are antiferromagnetic in Co$_2$V$_2$O$_7$. Moreover, by comparing the Curie-Weiss temperature $\theta_{CW}$ with the N\'{e}el temperature $T_N$, the frustration parameter \cite{BalentsNature} \emph{f} = $\mid \Theta_{CW} \mid$/$T_N$ is obtained to be 2.5(1), suggesting a moderate spin frustration in Co$_2$V$_2$O$_7$.
\begin{figure}[h]
	\centering
	\includegraphics[height=11.8cm]{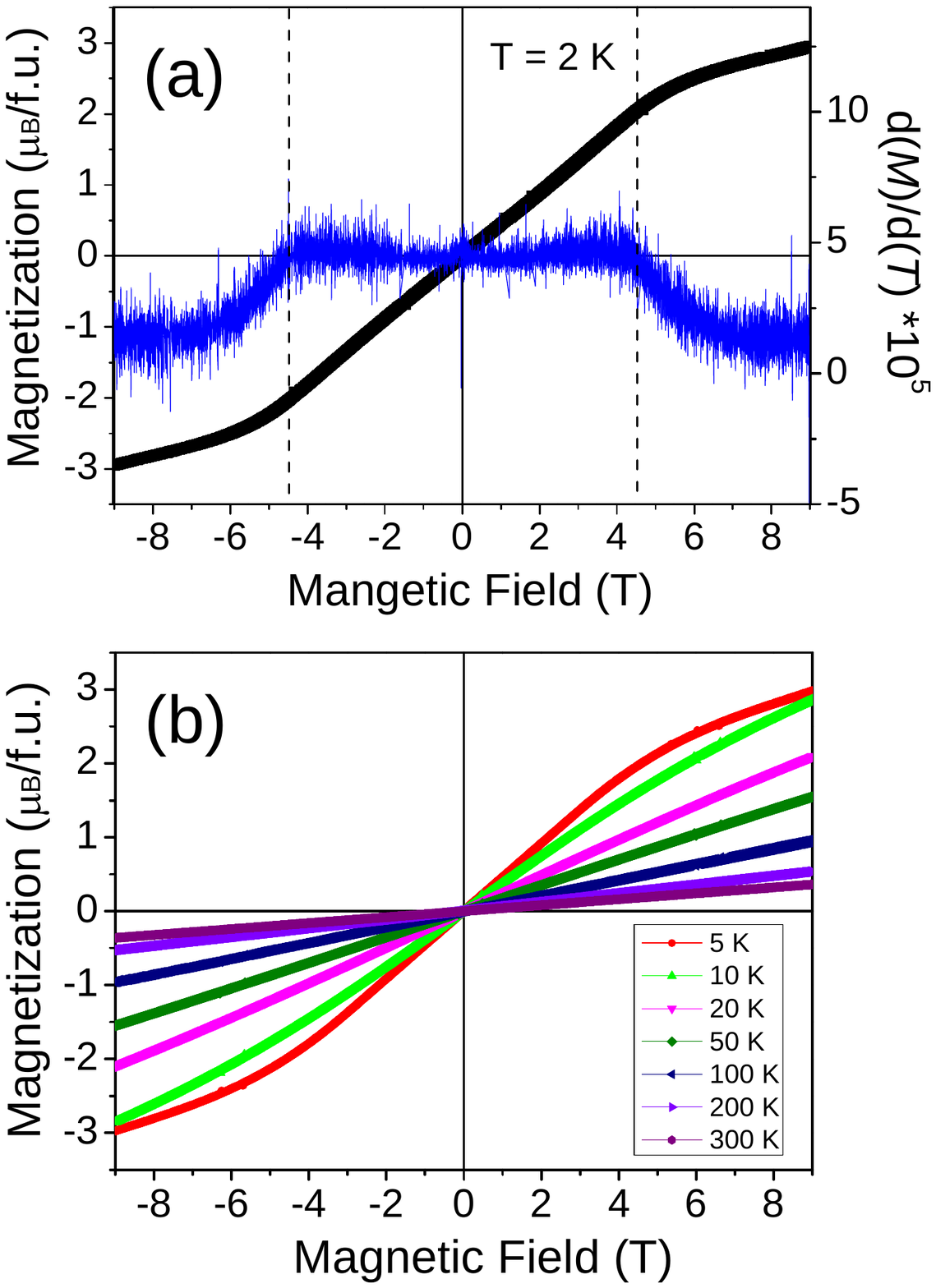}
	\caption{(a) Magnetization as a function of magnetic field for Co$_2$V$_2$O$_7$ at 2 K. The dashed line indicates the field-induced spin-flop transition at 4.2 T as determined from the anomaly in the derivative curve of the magnetization. (b) Magnetic field dependence of the magnetization of Co$_2$V$_2$O$_7$ measured at different temperatures between 5 and 300 K.}
	\label{fgr:example}
\end{figure}

Figure 3 (a) shows magnetization measured as a function of magnetic field at 2 K. Although the magnetization increase monotonically upon the application of magnetic field, the saturated magnetization still could not be reached upon the application of magnetic field up to 9 T. It is also noticed that a kink takes place at 4.2 T in \emph{M-H} curve, which suggestes that Co$_2$V$_2$O$_7$ undergoes a spin-flop transition with increasing magnetic field. The field-induced spin-flop transition is also clearly emphasized in the derivative curve of the magnetization data and it can be considered as the result of competition between Zeeman energy \emph{E} = -$\mu$$_0$$\vec{M}$$\cdot$$\vec{H}$ and magnetic anisotropy energy in  Co$_2$V$_2$O$_7$.
The \emph{M-H} curves at various temperatures are presented in Fig. 3(b). The spin-flop transition can still be observed at 5 K, while it is not discernible once temperature increased to above $T_N$. Nevertheless, considerable magnetization value, e.g. 2.85 $\mu_B/f.u.$ at 10 K and 9 T, can still be obtained above $T_N$, which indicates that the application of a magnetic field can not only change the ordering configuration but also enhances the magnetic ordering temperature of the Co$^{2+}$ moment. With further increase in temperature, the magnetization of Co$_2$V$_2$O$_7$ decreased and exhibited typical antiferromagnetic behavior.

\subsection{Crystal structure and non-collinear magnetic ground state determined by NPD}

\begin{figure*}[!t]
    \centering
    \includegraphics[height=8.5cm]{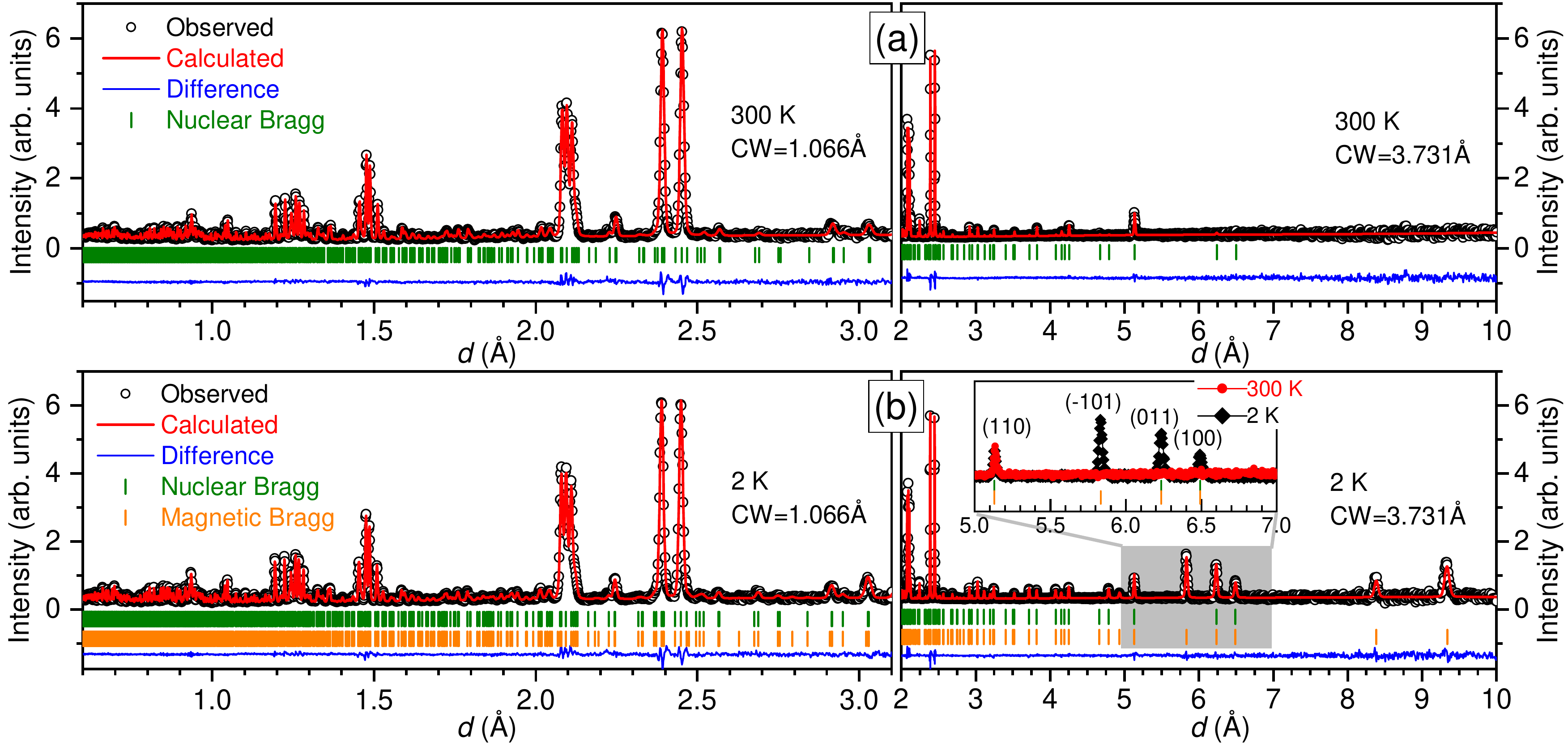}
    \caption{Rietveld refinement results of neutron powder diffraction patterns for Co$_2$V$_2$O$_7$ at (a) 300 K and (b) 2 K. The left and right panels show the data collected at the center-wavelength (CW) 1.066~{\AA}  and  3.731~{\AA}, respectively. The circles represent the observed intensity and the solid line is the calculated pattern. The difference between the observed and calculated intensities is shown at the bottom. The vertical bars indicate the expected nuclear and magnetic Bragg reflection positions. The inset shows a part of the diffraction pattern for 2 K and 300 K data indicating the additional magnetic reflections upon cooling.}
    \label{fgr:example2col}
\end{figure*}

In order to clarify the magnetic ground state of Co$_2$V$_2$O$_7$, neutron diffraction patterns were collected in a wide temperature range from 2 to 300 K. Fig. 4 shows the NPD patterns of Co$_2$V$_2$O$_7$ at 300 K and 2 K. Two different center wavelengths, i.e. 1.066 and 3.731 {\AA} are applied during the measurements so that a great number of diffraction peaks can be obtained in a large \emph{d}-space from 0.5 to 12 {\AA}. Two neutron diffraction patterns collected with two different center wavelengths were simultaneously refined for each temperature. At 300 K, Co$_2$V$_2$O$_7$ crystallizes in the monoclinic phase with space group $P 2_1/c$. By taking advantage of large neutron scattering cross section of Oxygen, the accurate position of Oxygen atom can be easily determined. However, the atomic position of Vanadium can not be determined precisely as the nuclei of Vanadium have a small neutron scattering cross section. Therefore, X-ray diffraction pattern are collected for  Co$_2$V$_2$O$_7$ at 300 K to determine the atomic position of Vanadium since Vanadium is a strong scatterer of X-rays. The detailed structural information for Co$_2$V$_2$O$_7$ at 300 K, as obtained from combining refinement of both NPD and XRD data, are given in Table I. The crystal structure of Co$_2$V$_2$O$_7$ and the octahedra of CoO$_6$ and are illustrated in Fig. 1. The bond length and bond angle of CoO$_6$ octahedra at 300 K is shown in Table II. The average Co-O bond length of CoO$_6$ for two nonequivalent Co sites, Co$_1$ and Co$_2$, are 2.074(5) and 2.083(5) {\AA} respectively, indicating a slightly different degree of distortion for these two octahedra. The crystal structure consists two non-equivalent Co sites. With decreasing temperature, the monoclinic structure of Co$_2$V$_2$O$_7$ persists down to 2 K accompanied by a shrinkage of lattice. The results of structural refinements at 2 K are also presented in Table I.

\begin{table*}
	\small
	\caption{\ Crystal structure of Co$_2$V$_2$O$_7$,Monoclinic Space group $p2_1/c$}
	\label{tbl:example}
	\begin{tabular*}{\textwidth}{@{\extracolsep{\fill}}lllllllllllll}
		\hline
		\multicolumn{2}{l}{Temperature} & \multicolumn{4}{c}{300 K} & & \multicolumn{2}{l}{Temperature} & \multicolumn{4}{c}{2 K}\\
		\cline{1-6}
		\cline{8-13}
		\multicolumn{2}{c}{\emph{a} = 6.5932(2) {\AA} } &    \multicolumn{2}{c}{ \emph{b} = 8.3775(2) {\AA}} & \multicolumn{2}{c}{   \emph{c} = 9.4782(2) {\AA} } & &
		\multicolumn{2}{c}{\emph{a} = 6.5860(2) {\AA} } &    \multicolumn{2}{c}{ \emph{b} = 8.3681(2) {\AA}} & \multicolumn{2}{c}{   \emph{c} = 9.4635(2) {\AA} } \\
		\multicolumn{2}{c}{$\alpha$ = 90$^\circ$ } &    \multicolumn{2}{c}{$\beta$ = 100.218(2)$^\circ$ } & \multicolumn{2}{c}{$\gamma$ = 90$^\circ$ } & &
		\multicolumn{2}{c}{$\alpha$ = 90$^\circ$ } &    \multicolumn{2}{c}{$\beta$ = 100.191(2)$^\circ$ } & \multicolumn{2}{c}{$\gamma$ = 90$^\circ$ } \\
		\cline{1-6}
		\cline{8-13}
		Type & Site & \emph{x} & \emph{y} & \emph{z} & B$_{iso}$  &  &    Type & Site & \emph{x} & \emph{y} & \emph{z} & B$_{iso}$  \\
		Co$_1$ & 4e &  0.148(3)   & 0.123(3)   &  0.463(2)  & 0.45(2) &  & Co$_1$ & 4e & 0.149(3)  & 0.121(3)  & 0.469(2)    &  0.24(1)   \\
		Co$_2$ & 4e &  0.310(3)   & 0.382(3)   &  0.685(2)  & 0.45(2) &  & Co$_2$ & 4e & 0.308(3)  & 0.389(3)  & 0.680(2)    &  0.24(1)   \\
		V$_1$  & 4e &  0.34(2)    & 0.75(2)    &  0.53(2)   & 0.45(2) &  & V$_1$  & 4e & 0.35(2)   & 0.75(2)   & 0.53(2)     &  0.24(1)   \\
		V$_2$  & 4e &  0.19(2)    & 0.01(2)    &  0.81(2)   & 0.45(2) &  & V$_2$  & 4e & 0.18(2)   & 0.01(2)   & 0.82(2)     &  0.24(1)   \\
		O$_1$  & 4e &  0.604(1)   & 0.130(1)   &  0.123(1)  & 0.55(1) &  & O$_1$  & 4e & 0.604(1)  & 0.130(1)  & 0.124(1)    &  0.19(1)   \\
		O$_2$  & 4e &  0.428(1)   & 0.126(1)   &  0.396(1)  & 0.49(1) &  & O$_2$  & 4e & 0.427(1)  & 0.127(1)  & 0.395(1)    &  0.07(1)   \\
		O$_3$  & 4e &  0.170(1)   & 0.369(1)   &  0.461(1)  & 0.55(2) &  & O$_3$  & 4e & 0.176(1)  & 0.370(1)  & 0.461(1)    &  0.16(1)   \\
		O$_4$  & 4e &  0.255(1)   & 0.360(1)   &  0.183(1)  & 0.37(1) &  & O$_4$  & 4e & 0.257(1)  & 0.359(1)  & 0.183(1)    &  0.04(1)   \\
		O$_5$  & 4e &  0.678(1)   & 0.373(1)   &  0.347(1)  & 0.61(2) &  & O$_5$  & 4e & 0.679(1)  & 0.373(1)  & 0.348(1)    &  0.19(1)   \\
		O$_6$  & 4e &  0.028(1)   & 0.081(1)   &  0.246(1)  & 0.48(1) &  & O$_6$  & 4e & 0.028(1)  & 0.083(1)  & 0.246(1)    &  0.12(1)   \\
		O$_7$  & 4e &  0.852(1)   & 0.378(1)   &  0.006(1)  & 0.52(1) &  & O$_7$  & 4e & 0.852(1)  & 0.379(1)  & 0.007(1)    &  0.21(1)   \\
		\cline{1-6}
		\cline{8-13}
		\multicolumn{2}{l}{300 K, 1.066{\AA}} & \multicolumn{4}{c}{ R$_p$ = 4.03 , R$_{wp}$ = 5.82 , $\chi$ = 0.50  } & & \multicolumn{2}{l}{2 K, 1.066{\AA} } & \multicolumn{4}{c}{ R$_p$ = 4.47, R$_{wp}$ = 6.32 , $\chi$ = 0.58 } \\
		\multicolumn{2}{l}{300 K, 3.731{\AA}} & \multicolumn{4}{c}{ R$_p$ = 3.72 , R$_{wp}$ = 5.35 , $\chi$ = 0.15  }& & \multicolumn{2}{l}{2 K, 3.731{\AA},} & \multicolumn{4}{c}{ R$_p$ = 5.39, R$_{wp}$ = 8.46 , $\chi$ = 0.39 } \\
		\hline
	\end{tabular*}
\end{table*}

\begin{table}[h]
	\small
	\caption{\ The bond length and bond angle of CoO$_6$ octahedra.}
	\label{tbl:example}
	\begin{tabular*}{0.48\textwidth}{@{\extracolsep{\fill}}llll}\hline
		Co$_1$&300 K&Co$_2$&300 K\\
		\hline
		O$_2$&2.0521 \AA&O$_1$&2.1298 \AA\\
		O$_3$&2.0675 \AA&O$_2$&2.0105 \AA\\
		O$_4$&2.0878 \AA&O$_3$&2.1642 \AA\\
		O$_6$&2.0956 \AA&O$_4$&2.0663 \AA\\
		O$_7$&2.0649 \AA&O$_5$&2.0677 \AA\\
		O$_7$&2.0746 \AA&O$_6$&2.0592 \AA\\
		O$_2$-Co$_1$-O$_3$&85.16 $^\circ$&O$_1$-Co$_2$-O$_4$&96.40 $^\circ$\\
		O$_2$-Co$_1$-O$_7$&94.63 $^\circ$&O$_1$-Co$_2$-O$_5$&87.08 $^\circ$\\
		O$_3$-Co$_1$-O$_7$&94.65 $^\circ$&O$_4$-Co$_2$-O$_6$&88.93 $^\circ$\\
		O$_7$-Co$_1$-O$_7$&86.44 $^\circ$&O$_5$-Co$_2$-O$_6$&87.67 $^\circ$\\
		O$_4$-Co$_1$-O$_6$&173.43 $^\circ$&O$_2$-Co$_2$-O$_3$&174.35 $^\circ$\\
		\hline	
		\multicolumn{2}{c}{Average Co-O bond length}\\
		300 K &Co$_1$-O =2.0737&Co$_2$-O =2.0829\\	
		\hline
	\end{tabular*}
\end{table}

In contrast to the NPD patterns at 300 K, extra peaks with considerable intensities show up once the sample was cooled to 2 K, which is below the N\'{e}el temperature \emph{T}$_N$ = 6.3 K. Most of new emerging peaks possess relatively large \emph{d}-value and can be assigned as magnetic Bragg peaks, indicating the formation of long range magnetic order of Co$^{2+}$ moments. It is also found that all magnetic reflections can be indexed as commensurate reflections with the magnetic propagation wave vector \textbf{k} = (0, 0, 0), which means that the magnetic structure is commensurate in nature.
\begin{table}[h]
	\small
	\caption{\ Nonzero basis vectors (BV) of the irreducible representations (IR) and positional coordinates for Co atoms that was used to describe the non-collinear magnetic structure of Co$_2$V$_2$O$_7$ with space group $P2_1/c$ and propagation vector \emph{k} = (0, 0, 0).}
	\label{tbl:example}
	\begin{tabular*}{0.48\textwidth}{@{\extracolsep{\fill}}lllcccccc}
		\hline
		& & & \multicolumn{3}{c}{BV Components} & \multicolumn{3}{c}{Positional coordinates} \\
		\cline{4-6}
		\cline{7-9}
		IR& BV & Atom & \emph{m}$_{\shortparallel a}$ & \emph{m}$_{\shortparallel b}$ & \emph{m}$_{\shortparallel c}$ & \emph{x} & \emph{y} & \emph{z}\\
		\hline
		$\Gamma$$_2$  & $\psi$$_4$ & 1 & 1 & 0 & 0 &   $x$ & $y$ & $z$\\
		&            & 2 & -1 & 0 & 0 &  $\bar{x}$+1 & $y$+$\frac{1}{2}$ & $\bar{z}$+$\frac{1}{2}$ \\
		&            & 3 & -1 & 0 & 0 &  $\bar{x}$+1 & $\bar{y}$+1 & $\bar{z}$+1 \\
		&            & 4 & 1 & 0 & 0 &   $x$ & $\bar{y}$+$\frac{1}{2}$ & $z$+$\frac{1}{2}$ \\
		& $\psi$$_5$ & 1 & 0 & 1 & 0 &   $x$ & $y$ & $z$\\
		&            & 2 & 0 & 1 & 0 &   $\bar{x}$+1 & $y$+$\frac{1}{2}$ & $\bar{z}$+$\frac{1}{2}$ \\
		&            & 3 & 0 & -1 & 0 &  $\bar{x}$+1 & $\bar{y}$+1 & $\bar{z}$+1 \\
		&            & 4 & 0 & -1 & 0 &  $x$ & $\bar{y}$+$\frac{1}{2}$ & $z$+$\frac{1}{2}$ \\
		& $\psi$$_6$ & 1 & 0 & 0 & 1 &   $x$ & $y$ & $z$\\
		&            & 2 & 0 & 0 & -1 &  $\bar{x}$+1 & $y$+$\frac{1}{2}$ & $\bar{z}$+$\frac{1}{2}$ \\
		&            & 3 & 0 & 0 & -1 &  $\bar{x}$+1 & $\bar{y}$+1 & $\bar{z}$+1 \\
		&            & 4 & 0 & 0 & 1 &   $x$ & $\bar{y}$+$\frac{1}{2}$ & $z$+$\frac{1}{2}$ \\
		\hline
	\end{tabular*}
\end{table}

Symmetry analysis allows the determination of the symmetry-allowed magnetic structures based on the representation theory. Magnetic symmetry analysis was carried out for Co$_2$V$_2$O$_7$ using the program SARAh \cite{wills2000new} by which the allowed symmetry couplings in the form of irreducible representation and their basis vectors can be deduced. It is known that two Co$^{2+}$ ions occupy two inequivalent crystallographic Wyckoff sites 2\emph{e} in Co$_2$V$_2$O$_7$ with the crystallographic space group \emph{P2$_1$/c}. With the commensurate magnetic propagation vector \textbf{k} = (0, 0, 0), both Co ions are found to have four one-dimensional irreducible magnetic representations (IRs), i.e. $\Gamma$$_1$, $\Gamma$$_2$, $\Gamma$$_3$, $\Gamma$$_4$, and each of them is composed of three basis vectors (BV) with only real components. Assuming that Co atoms located at two nonequivalent crystallographic sites carry magentic moments with different amplitude as in Co$_3$V$_2$O$_8$ \cite{ChenPRB} counterpart, all possible IRs are adopted into the magnetic structural model and used to fit against the NPD pattern at 2 K.

As demonstrated in Fig. 4(d), a satisfactory fitting of all magnetic reflections can only be obtained by using IR $\Gamma$$_2$ with Shubnikov space group of \emph{P2$_1$/c'}, while other three representations can be ruled out due to poor fit of magnetic reflections. The decomposition of the magnetic representation $\Gamma$$_2$ in terms of basis vectors are given in Table III.  According to $\Gamma$$_2$, the Co moments align antiferromagnetically in the magnetic unit cell and the refined vector components of magnetic moment are obtained to be different for two individual Co 2\emph{e} sites, i.e. $m_x$ = -0.26(5) $\mu_B$, $m_y$ = 1.77(5) $\mu_B$ and $m_z$ = 0.97 (5) $\mu_B$ for Co$_1$ site, $m_x$ = 0.35(5) $\mu_B$, $m_y$ = 2.38(5) $\mu_B$ and $m_z$ = 1.27 (5) $\mu_B$ for Co$_2$ site, respectively. It is obvious that the canted Co moments ordered antiferromagnetically in \emph{b-c} plane and form a spin-chain-like structure along \emph{c}-axis, as illustrated in Fig. 5. The refined magnetic moments of Co in the spin chain are 2.06 (5) and 2.69 (5) $\mu_B$ for Co$_1$ and Co$_2$, respectively. The canted angle of Co moment toward \emph{c}-axis is around 26(1)$^\circ$. Nevertheless, the tilt of Co spin does not follow the tilt of CoO$_6$ octahedra, suggesting the negligible influence of electric multipolar field generated by oxygen neighbors on the orbitals of Co ions. The antiferromagnetic ground state of Co$_2$V$_2$O$_7$ is apparently determined by the interplay between spin-spin exchange interaction and magnetocrystalline anisotropy, which are largely dependent on the electronic configuration of Co ions. It is known that the up–up–down–down spin arrangement along the alternating array of two magnetic sites can break the inversion symmetry, and the inequivalent exchange striction working between the up–up (or down–down) spin pair and the up–down (or down–up) can produce electric polarization along the chain direction \cite{Tokura_2014}. Obviously, the exchange-striction mechanism can be applied to Co$_2$V$_2$O$_7$ with canted $\uparrow$$\uparrow$$\downarrow$$\downarrow$ spin chains (see Fig. 5) and the electric polarization will emerge as a consequence.

\begin{figure}[h]
	\centering
	\includegraphics[height=7.5cm]{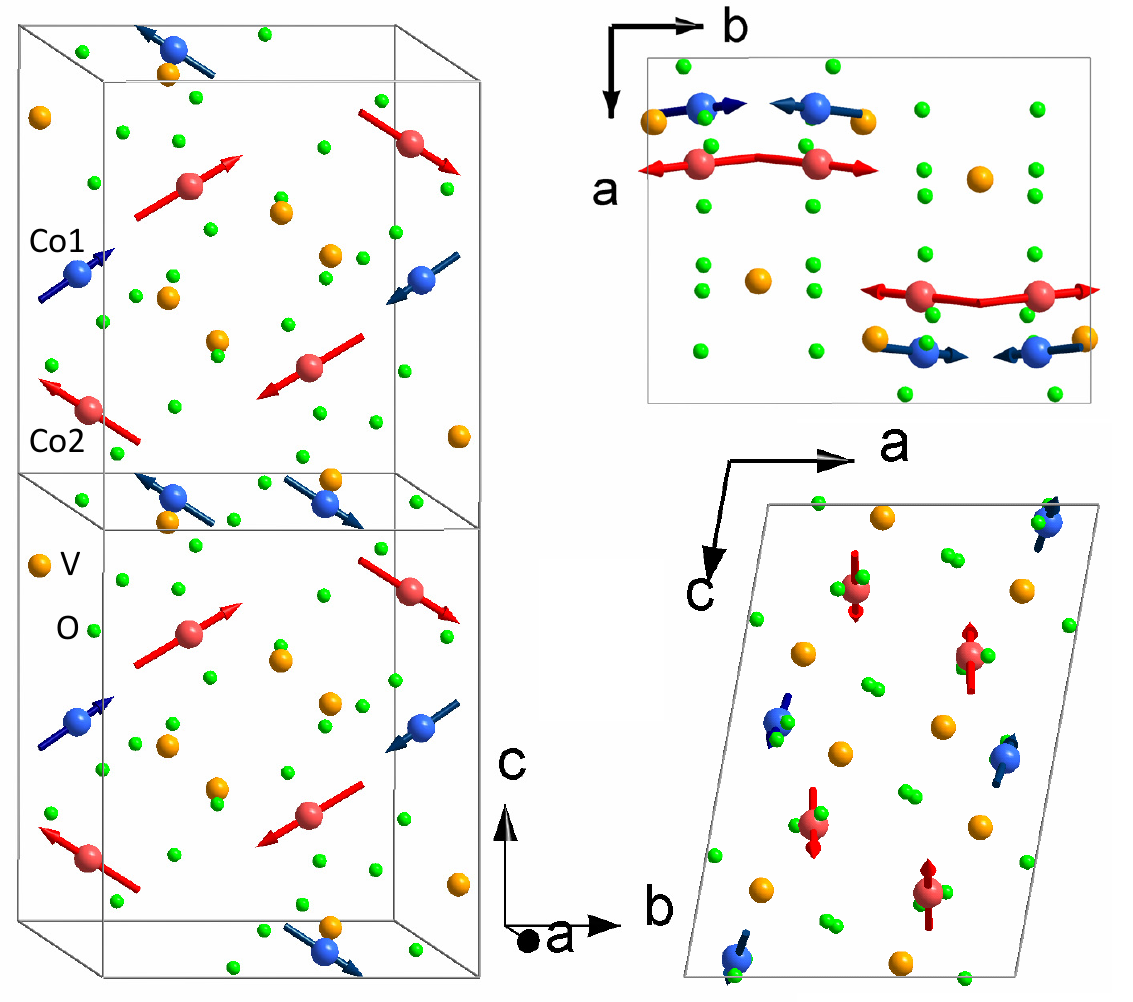}
	\caption{Illustration of the non-collinear magnetic structure of Co$_2$V$_2$O$_7$ below the magnetic transition temperature. The Co moments ordered antiferromagnetically in b-c plane. The magnetic unit cell is identical to the monoclinic structural unit cell which is outlined with the gray lines.}
	\label{fgr:example}
\end{figure}

\subsection{Calculation results on magnetic ground state}
The calculation started with the experimental spin structure to investigate the electronic structure and magnetic properties of Co$_2$V$_2$O$_7$. Two sets of non-collinear calculations are performed: One was with a constraint that fixes Co magnetic moment directions during self-consistent calculations, and the other was without this constraint so that both the directions and magnitudes of each Co magnetic moment were relaxed.
Table \uppercase\expandafter{\romannumeral4} shows the magnetic moment components of each Co atom obtained from NC magnetic calculations with and without relaxing the spin directions. Experimental values are also included to compare. The calculations were carried out using DFT+\emph{U} with \emph{U} = 3.2 eV. The calculations show similar magnetic moments for two non-equivalent Co sites while the experiment found noticeably different magnetic moments between two Co sites. The further relaxation of spin direction only introduces small changes on the amplitudes and directions of Co moments, suggesting the obtained spin structure is rather stable.

\begin{table}[h]
	\small
	\caption{\ Comparison of measured and calculated magnetic moments of two Co sites in Co$_2$V$_2$O$_7$. The Co magnetic moments are in the unit of $\mu_B$.}
	\label{tbl:example}
	\begin{tabular*}{0.48\textwidth}{@{\extracolsep{\fill}}lrrrrrrrrr}\hline
		& \multicolumn{3}{c}{Experiment}& \multicolumn{3}{c}{Constrained}& \multicolumn{3}{c}{Relaxed}\\
		\cline{2-4}
		\cline{5-7}
		\cline{8-10}
		&m$_x$&m$_y$&m$_z$&m$_x$&m$_y$&m$_z$&m$_x$&m$_y$&m$_z$\\[1mm]
		\hline
		Co$_1$ & -0.260 & 1.773 & 0.968 & -0.34 & 2.32 & 1.27 & -0.30 & 2.33 & 1.25 \\
		&0.260&1.773&-0.968&0.34&2.32&-1.27&0.30&2.33&-1.25\\
		&0.260&-1.773&-0.968&0.34&-2.32&-1.27&0.29&-2.34&-1.25\\
		&-0.260&-1.773&0.968&-0.34&-2.32&1.27&-0.30&-2.34&1.24\\
		Co$_2$&0.348&2.376&1.271&0.34&2.33&1.24&0.37&2.37&1.15\\
		&-0.348&-2.376&-1.271&-0.34&-2.33&-1.24&-0.37&-2.38&-1.14\\
		&-0.348&2.376&-1.271&-0.34&2.33&-1.24&-0.36&2.37&-1.15\\
		&0.348&-2.376&1.271&0.34&-2.33&1.24&0.37&-2.37&1.15\\
		\hline	
	\end{tabular*}
\end{table}

Atom-projected partial densities of states (PDOS) of the relaxed spin configuration are shown in Fig. 6. The calculations were performed within GGA and GGA+\emph{U} with \emph{U} = 3.2 eV, respectively. In the case of \emph{U} = 0 eV, there are sharp Co-DOS peaks near the Fermi energy. The bandgap sits between two sharp DOS peaks, which mostly consist of Co-\emph{d} states that barely hybridize with other states and form  flat bands. The bandgap is about 0.2 eV within GGA and depends on the applied \emph{U} parameters with GGA+\emph{U}. The PDOS with \emph{U} = 3.2 eV are shown in Fig. 6 (b). As expected, a larger bandgap is obtained. It seems that \emph{U} potential is able to rearrange Co states to promote the hybridization of Co-\emph{d} states with other states so that NC magnetic structure becomes stable. It would be interesting to measure the bandgap, which may guide justifying the \emph{U} parameter.

\begin{figure}[h]
	\centering
	\includegraphics[height=10cm]{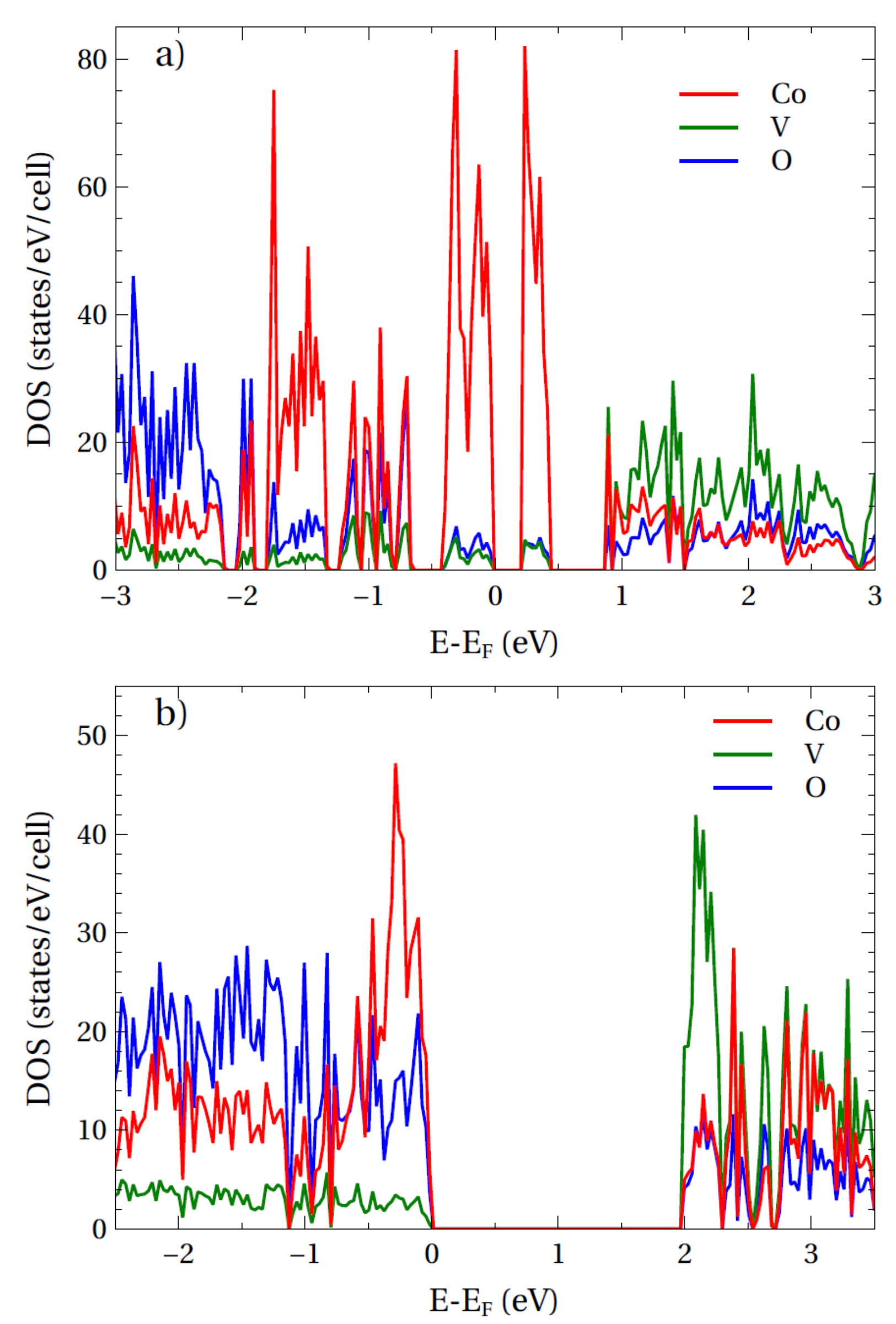}
	\caption{Atom-projected partial densities of states calculated within (a) GGA and (b) GGA+\emph{U} with \emph{U} = 3.2 eV.}
	\label{fgr:example}
\end{figure}

 Considering that the relaxed spin configuration is not necessary the ground state spin configuration, here, we calculate and compare the total energies of FM, AFM, and NC magnetic structures using VASP. The dependence of these energy differences on the correlation parameter U is also investigated. The plain DFT calculations show that FM state has the lowest energy, which is 3.86 eV/Co and 19.88 eV/Co lower than the AFM state and the NC state, respectively. It is well known that LDA (GGA)+\emph{U} method gives better agreement with experimental results than plain LDA (GGA) method for oxide gap materials because of strong correlation. We performed calculations with three different \emph{U} parameters to understand their effects on the magnetic states. Table \uppercase\expandafter{\romannumeral5} shows the total energies for FM, AFM and NC magnetic structures with various \emph{U} values. It shows that the magnetic ground state depends on the magnitude of \emph{U} potential. A larger \emph{U} parameter tends to stabilize the NC state.

\begin{table}[h]
	\small
	\caption{\ Calculated total energy (meV/Co) of FM, AFM, and experimental non-collinear (denoted as NC) magnetic states with \emph{U} = 0, 1.6, 3.2, and 4.8 eV using VASP.}
	\label{tbl:example}
	\begin{tabular*}{0.48\textwidth}{@{\extracolsep{\fill}}lrrrrr}\hline
		\emph{U} (eV)&0&1.6&3.2&4.8\\
		\hline
		FM&0.00&0.00&0.00&0.00\\
		AFM&3.86&-8.59&20.78&18.72\\
		NC&19.88&-7.17&-5.06&-3.18\\
		\hline
		
	\end{tabular*}
\end{table}

\subsection{Anisotropic magnetoelastic coupling and thermal expansion}
The change of individual lattice parameters with temperature can be obtained through the refinement of NPD patterns collected at different temperatures. As shown in Fig. 7 (a)-(d), the onset of magnetoelastic distortions below \emph{T}$_N$ = 6.3 K is clearly visible for lattice parameters \emph{a} and \emph{b}, while the lattice parameter \emph{c} and $\beta$ stay almost constant upon crossing the N\'{e}el temperature. It is worth noting that the sign of the deviation for \emph{a} is opposite to that for \emph{b}, indicating the existence of anisotropic magnetoelastic coupling in Co$_2$V$_2$O$_7$. The corresponding unit-cell volume is shown in Fig. 7(e). Although \emph{a} elongates and \emph{b} shrinks upon cooling below T$_N$, they compensate each other in the volumetric expansion. The unit-cell volume \emph{V} shows no anomamalous change and increases monotonically with temperature.
\begin{figure}[h]
	\centering
	\includegraphics[height=10cm]{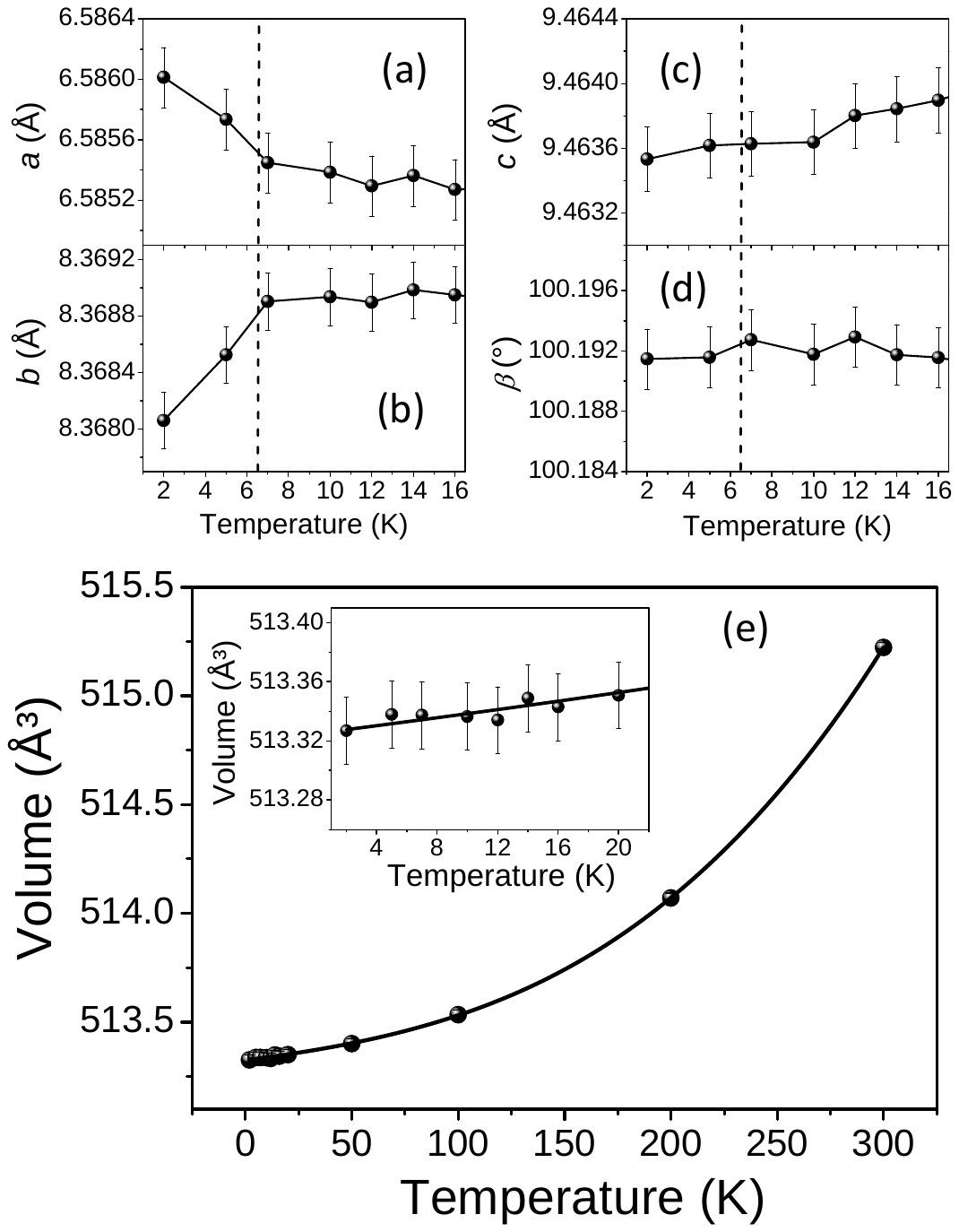}
	\caption{(a)-(d) Temperature dependence of the estimated lattice parameters \emph{a}, \emph{b} \emph{c} and $\beta$. (e) Temperature dependence of the volume of unit-cell of Co$_2$V$_2$O$_7$. The inset shows the variation of unit-cell volume in low temperature range.}
	\label{fgr:example}
\end{figure}

Because the tilt of Co spin does not follow the tilt of CoO$_6$ octahedra, we speculate that the influence of electric multipolar field generated by oxygen neighbors on the orbitals of Co ions is negligible. The antiferromagnetic ground state of Co$_2$V$_2$O$_7$ is apparently determined by the interplay between spin-spin exchange interaction and single-ion anisotropy. In order to understand anisotropic magnetoelastic coupling and thermal expansion, we correlate the variation of distance between different Co ions in same spin-chain along the \emph{c}-axis to antiferromagnetic phase transition. The change of distance between neighbouring Co sites at different temperatures is shown in Fig. 8(a). Both of the distance of Co$_1$-Co$_1$ and Co$_2$-Co$_2$ decrease with the increasing of temperature and remain almost constant above $T_N$. Figure 8(b) display the temperature dependence of the change of distance between different Co sites. According to the non-colliner magnetic structure illustrated in Fig. 5, two non-equivalent Co sites in same spin-chain form parallel or antiparallel spin configurations. The distance of Co$_1$-Co$_2$ with parallel spin ordering shows same behavior as the lattice parameter \emph{b}, and the distance of Co$_1$-Co$_2$ with antiparallel spin ordering shows same behavior as the lattice parameter \emph{a}. This is clearly evidence that the anisotropic magnetoelastic coupling and thermal expansion of lattice parameters are most likely associated with the different exchange interactions between two neighboring Co spins.
\begin{figure}[h]
	\centering
	\includegraphics[height=12cm]{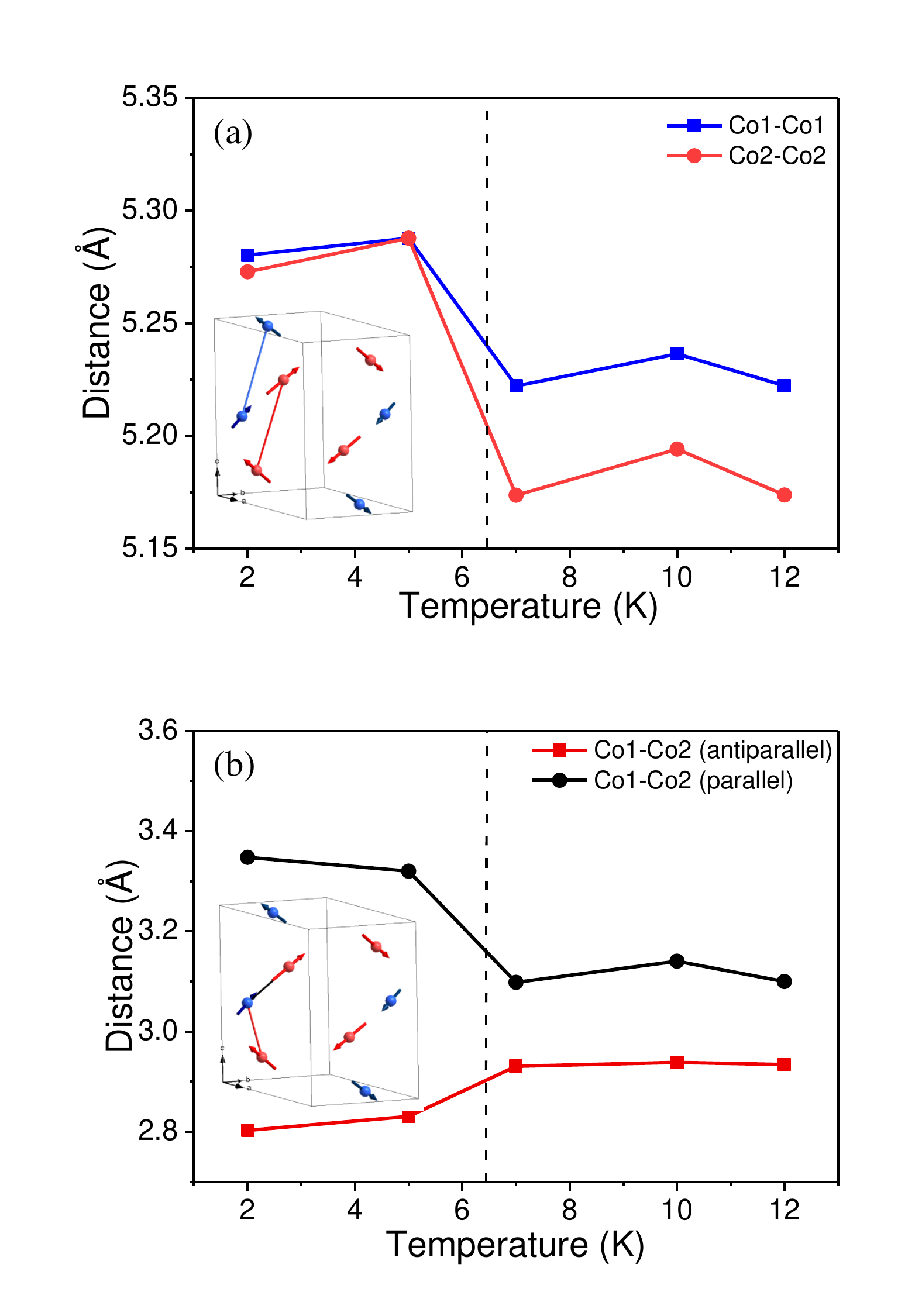}
	\caption{Temperature dependence of the change of distance between different Co ions in same spin-chain along \emph{c}-axis.}
	\label{fgr:example}
\end{figure}

\section{conclusions}
In summary, we have studied the magnetic properties of Co$_2$V$_2$O$_7$ by extensive approaches, such as magnetization measurements, neutron powder diffraction measurements and theoretical calculations. By modeling the temperature dependence of the magnetization with the Curie-Weiss law, we find that the Curie-Weiss temperature is negative, e.g., -16.1(5) K, indicating dominating antiferromagnetic coupling. A spin-flop transition was observed while magnetic field was applied below $T_N$= 6.3 K. By applying neutron powder diffraction methods, we confirm that Co$_2$V$_2$O$_7$ remains monoclinic from 300 K down to 2 K. The magnetic ground state was determined unambiguously based on NPD results by using representation analysis combined with magnetic Rietveld refinements. Two non-equivalent Co sites with different magnetic moments form a spin-chain-like structure along \emph{c}-axis and canted antiferromagnetic ordering in \emph{b-c} plane. By comparing total energies between different magnetic states, we found that the theoretical magnetic ground state is sensitive to the \emph{U} parameter and GGA+\emph{U} calculations show that the non-collinear magnetic state possesses a lower energy than FM and AFM states. The ferroelectricity in Co$_2$V$_2$O$_7$ can be understood based on exchange striction model, which can be applied to canted $\uparrow$$\uparrow$$\downarrow$$\downarrow$ spin configurations. Moreover, the temperature dependence of lattice parameters in Co$_2$V$_2$O$_7$ exhibits an anisotropic magnetoelastic coupling between the \emph{a}, \emph{b} and \emph{c} below $T_N$.

\section{\label{sec:level1}Acknowledgment}
	This work was supported by the National Natural Science Foundation of China (Grant Nos. 11874023) and Project Based Personnel Exchange Program (PPP) with China Scholarship Council (CSC Nos. 2016-6041) and German Academic Exchange Service (Project-ID 57219934). Part of the research conducted at SNS was sponsored by the Scientific User Facilities Division, Office of Basic Energy Sciences, US Department of Energy. WHJ would like to acknowledge financial support from the China Scholarship Council(CSC).

\appendix

\bibliography{prb}

\end{document}